\documentclass[sigconf]{acmart}
\AtBeginDocument{%
  }
\setcopyright{acmlicensed}
\copyrightyear{2018}
\acmYear{2018}
\acmDOI{XXXXXXX.XXXXXXX}
\acmConference[Submitted to CIKM ’25]%
  {Proceedings of the 34th ACM International Conference on Information and Knowledge Management (CIKM '25)}%
  {November 10--14, 2025}%
  {COEX, SEOUL, KOREA}
\acmBooktitle{Proceedings of the 34th ACM International Conference on Information and Knowledge Management (Submitted to CIKM '25), November 10--14, 2025, COEX, SEOUL, KOREA}
\acmISBN{978-1-4503-XXXX-X/2018/06}

\usepackage{graphicx}
\usepackage{subcaption}
\usepackage{caption}
\usepackage{float}
\usepackage{enumitem}
\usepackage{multirow}
\usepackage{makecell} 
\usepackage{titlesec}
\usepackage{balance}
\usepackage{bm}
\titlespacing{\section}{0pt}{2ex}{1ex}
\titlespacing{\subsection}{0pt}{2ex}{1ex}
\titlespacing{\subsubsection}{0pt}{1ex}{1ex}

\begin{document}
\title{SSEmb: A Joint Structural and Semantic Embedding Framework for Mathematical Formula Retrieval}

\author{Ruyin Li}
\affiliation{%
  \institution{School of Mathematical Sciences,\\Beihang University}
  \city{Beijing}
  \country{China}
}
\email{liruyin@buaa.edu.cn}
\author{Xiaoyu Chen}
\authornote{Corresponding Author}
\affiliation{%
  \institution{School of Mathematical Sciences,\\Beihang University}
  \city{Beijing}
  \country{China}
}
\email{chenxiaoyu@buaa.edu.cn}
\renewcommand{\shortauthors}{Ruyin Li and Xiaoyu Chen}

\begin{abstract}\noindent
Formula retrieval is an important topic in Mathematical Information Retrieval. We propose SSEmb, a novel embedding framework capable of capturing both structural and semantic features of mathematical formulas. Structurally, we employ Graph Contrastive Learning to encode formulas represented as Operator Graphs. To enhance structural diversity while preserving mathematical validity of these formula graphs, we introduce a novel graph data augmentation approach through a substitution strategy. Semantically, we utilize Sentence-BERT to encode the surrounding text of formulas. Finally, for each query and its candidates, structural and semantic similarities are calculated separately and then fused through a weighted scheme. In the ARQMath-3 formula retrieval task, SSEmb outperforms existing embedding-based methods by over 5 percentage points on P'@10 and nDCG'@10. Furthermore, SSEmb enhances the performance of all runs of other methods and achieves state-of-the-art results when combined with Approach0.
\end{abstract}

\maketitle

\section{Introduction}\noindent
In the contextualized ARQMath-3 formula retrieval task \cite{10.1007/978-3-031-13643-6_20}, a formula from a Community Question Answering (CQA) post is used as a query to retrieve related formulas from a corpus of posts. This task is fundamental to Mathematical Information Retrieval (MIR), as it helps users explore the meaning, derivation, and applications of a formula, supporting downstream tasks such as problem solving, concept learning, and mathematical reasoning in the cross domain. However, formulas with highly similar structures may belong to different domains with distinct meanings, while structurally dissimilar ones can contain semantically relevant terms, offering useful mathematical insights. Therefore, a key challenge for formula retrieval lies in retrieving mathematically meaningful information, which requires assessing relevance based not only on structural similarity, but also on the semantic cues provided by the surrounding text.
\begin{figure}[h]
  \centering
  \begin{subfigure}{0.33\linewidth}
    \includegraphics[width=\linewidth]{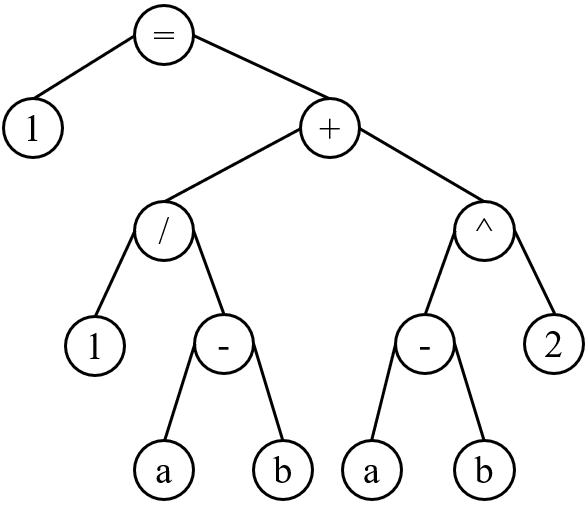}
    \captionsetup{skip=2pt}
    \caption{OPT}
    \label{Pic1a}
  \end{subfigure}
  \hspace{0.1\linewidth}
  \begin{subfigure}{0.3\linewidth}
    \includegraphics[width=\linewidth]{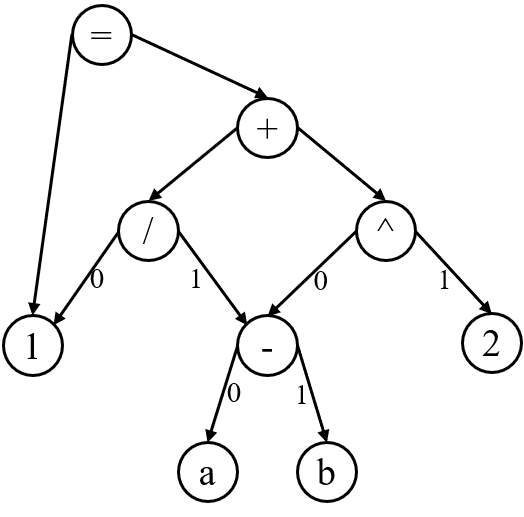}
    \captionsetup{skip=2pt}
    \caption{OPG}
    \label{Pic1b}
  \end{subfigure}
  \captionsetup{skip=6pt}
  \caption{Structural representations of $\bm{\frac{1}{a - b} + (a - b)^2 = 1}$.}
  \label{Pic1}
\end{figure}

Early research in this field mainly focused on matching formulas based on their text-based \cite{miller2003technical, misutka2008extending, kumar2012structure} or tree-based representations \cite{zhong2019structural, davila2016tangent3, davila2017layout, Wikimirs2013, WikiMirs3.02015, MCAT2016}, which often depend on hand-crafted matching rules and suffer from high computational costs. With the advancement of deep learning, embedding-based methods have emerged as a promising alternative. By encoding formulas to vectors, these methods allow efficient similarity computation in continuous space. Formula embedding techniques have evolved from symbol-level \cite{gao2017preliminaryexplorationformulaembedding, EquationEmbeddings2018, TopicEq2019, MathAMR2022} and tuple-level embedding \cite{Thanda2016, Tangent-CFT2019, Dai2020} to graph-level embedding \cite{Song2021}, which offers richer structural features. Specifically, Gao et al. \cite{gao2017preliminaryexplorationformulaembedding} proposed Symbol2Vec to learn symbol-level embeddings, and then Formula2Vec was introduced to learn distributed representations of formulas. Mansouri et al. \cite{Tangent-CFT2019} proposed Tangent-CFT, which performed a depth-first search over the Symbol Layout Tree (SLT) and Operator Tree (OPT) to linearize each formula to a tuple sequence. Each tuple was considered as a word, and then the FastText n-gram embedding model was applied to embed the formula. Mansouri et al. \cite{MathAMR2022} introduced Math Abstract Meaning Representation (MathAMR) graph by unifying Abstract Meaning Representation (AMR) graph and OPT to capture the meaning of a formula in one sentence, and then the linearized MathAMR graph was embedded with Sentence-BERT. Wang et al. \cite{GCL-MIR} used general Graph Contrastive Learning (GCL) to embed formulas represented as SLT and OPT. Song et al. \cite{Song2021} introduced the Operator Graph (OPG) representation by sharing identical subtrees within the OPT, characterizing structural features in a compact form, as depicted in Fig. \ref{Pic1}. Then Graph Neural Network (GNN) was employed to learn OPG embeddings. Generating high-quality graph-level embeddings shows promising potential and deserves further exploration. Furthermore, existing embedding-based methods do not fully utilize contextual semantic information. Effective integration of both structural and semantic cues remains an open challenge.

To address this, we propose {\bfseries SSEmb}, a novel embedding framework that jointly models structural features and contextual semantics of formulas. The framework consists of two core modules: Structural Embedding ({\bfseries StructEmb}) and Semantic Embedding ({\bfseries SemEmb}).

The {\bfseries StructEmb} module generates graph-level embeddings by applying Graph Contrastive Learning (GCL) \cite{GraphCL2020} on OPG. Accounting for the hierarchical structure of a formula, we introduce a graph data augmentation
approach through substructure substitution to enhance robustness and capture structural features from global to local levels.

The {\bfseries SemEmb} module leverages Sentence-BERT \cite{Sentence-BERT2019} to encode long surrounding text, in contrast that MathAMR \cite{MathAMR2022} only considers the sentence where the formula appears in. This enables the extraction of latent knowledge, such as usage scenarios, definitions, and domain-specific semantics associated with the formula.  

In summary, our contributions are threefold as follows:
\begin{itemize}[leftmargin=9pt, nosep]
\item We propose SSEmb, a joint embedding framework that captures both structural and semantic features of formulas.
\item We design a novel formula augmentation approach to improve the  expressiveness of structural embeddings.
\item We demonstrate that SSEmb outperforms existing embedding-based methods on the ARQMath-3 formula retrieval task and further enhances state-of-the-art approaches when combined.
\end{itemize}

\section{Methodology}
\begin{figure*}[t]
  \centering
  \includegraphics[width=\textwidth]{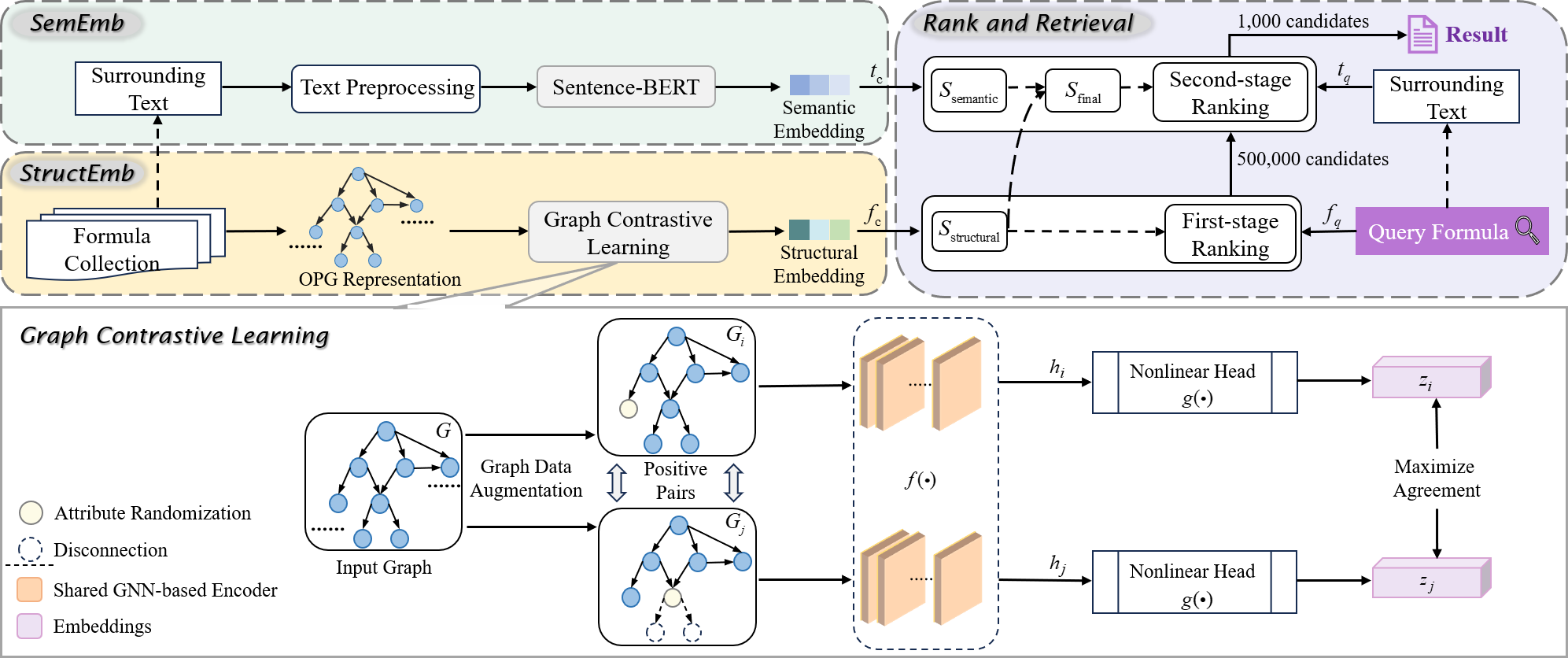}
  \captionsetup{skip=6pt}
  \caption{The SSEmb framework.}
  \label{Pic2}
\end{figure*}

\noindent
In this section, we provide a detailed description on SSEmb. Fig. \ref{Pic2} illustrates the overall framework, which consists of three modules: StructEmb, SemEmb, Rank and Retrieval.
\subsection{StructEmb}\noindent
The StructEmb module first converts formulas to OPG representations, and then leverages GCL to encode them. To align with the hierarchical nature of formulas, we design a graph data augmentation approach that introduces structural variations while preserving mathematical validity.
\subsubsection{OPG Representation.}
We represent a formula with OPG \cite{Song2021}, a labeled directed acyclic graph built on OPT. By sharing identical substructures, OPG provides a more compact representation of formula structures. Compared to OPT, OPG offers two main advantages: (1) It captures hierarchical structures with higher information density and expressive power; (2) Substructure sharing enables more efficient and effective graph data augmentation.
\subsubsection{Graph Contrastive Learning.}
In order to capture the structural features of formulas, we follow the general framework of GCL \cite{GraphCL2020} with modifications to specific components to generate formula embeddings. GCL performs training by minimizing contrastive loss in the embedding space to maximize the agreement between two augmented views of the same graph. As shown in Fig. \ref{Pic2}, the process mainly consists of the following four steps:

\textbf{(1) Graph data augmentation.}
Given a graph $G$, two related views $G_{i}$ and $G_{j}$ need to be generated as positive pairs through graph data augmentation. For formula embedding, it is crucial to select appropriate augmentations. StructEmb adopts two strategies: attribute masking \cite{GraphCL2020} and substructure substitution, which perturb node attributes and graph structures, respectively.

Conventional structure augmentations such as node dropping and edge perturbation often destroy formula integrity, leading to syntactical invalidity or semantical meaninglessness that hinder effective learning. To address this, we propose a substructure substitution strategy inspired by the hierarchical theory of formulas, as discussed in the WikiMirs system \cite{Wikimirs2013}. Higher-level structures capture the core computational skeleton of a formula, whereas lower-level nodes (e.g., variables or constants) have limited influence on the global structure. 

\begin{figure}[H]
  \centering
  \includegraphics[width=\linewidth]{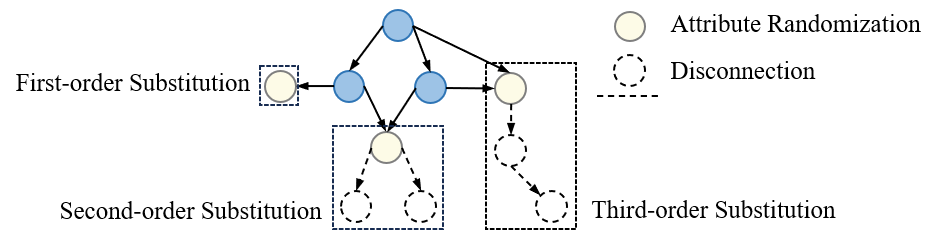}
  \captionsetup{skip=6pt}
  \caption{Substructure substitution on the OPG of a formula.}
  \label{Pic3}
\end{figure}

Substitution starts from the leaf nodes in the OPG representation and progressively selects higher-level substructures with predefined probabilities, by replacing the root node attribute of each selected substructure with a randomly generated wildcard and disconnecting the root node from lower-level nodes. Taking Fig. \ref{Pic3} as an example, the first-order substitution replaces leaf nodes (with probability $p_{1}$) to obscure specific variable information while preserving the overall structure and computation logic; the second-order substitution targets the parent nodes of leaf nodes (with probability $p_{2}$), introducing local structural variation to improve robustness; the third-order substitution replaces grandparent-level substructures (with probability $p_{3}$), encouraging the model to learn more global representations. To maintain mathematical coherence of augmented samples, higher-order substitutions are applied less frequently.

The computational complexity of substructure substitution mainly depends on the operations of locating and replacing. Locating leaf nodes and their ancestors in OPG requires traversing parent-child relationships, with a complexity of $O(\left | E \right |)$, where $\left | E \right |$ is the number of edges. Replacing involves modifying the selected substructures, with complexity $O(\left | V_{t} \right |)$, where $\left | V_{t} \right |$ is the number of affected nodes. The total complexity $O(\left | E \right |+\left | V_{t} \right |)$ remains comparable to conventional augmentation methods.

\textbf{(2) GNN-based encoder.}
We employ a multi-layer GNN-based encoder $f(\cdot )$ to extract graph-level representation vectors $h_{i}$ and $h_{j}$ for the augmented graphs $G_{i}$ and $G_{j}$, respectively . Each GNN layer adopts the Graph Isomorphism Network (GIN).

\textbf{(3) Nonlinear head.}
Nonlinear transformation $g(\cdot )$ is adopted after GNN layer to map the augmented representations to another latent space where the contrastive loss is calculated to obtain $z_{i}$ and $z_{j}$. The transformation is normalized and then ReLU is applied.

\textbf{(4) Contrastive loss function.}
A contrastive loss function $L(\cdot )$ is defined to enforce maximizing the consistency between positive pairs $z_{i}$ and $z_{j}$ compared with negative pairs. Here $L(\cdot )$ adopts InfoNCE Loss \cite{InfoNCELoss2018}.
\subsection{SemEmb}\noindent
The SemEmb module captures contextual semantics of each formula by extracting and preprocessing the surrounding text, followed by encoding with a pre-trained Sentence-BERT model.

\subsubsection{Text Extracting and Preprocessing.}
The accompanying text is extracted from the post where a formula is located, which often reveals crucial information about the formula’s latent knowledge, such as the research domain, usage, definitions, and underlying concepts. To improve calculation efficiency and avoid redundant information interference, the string length of the extracted text is truncated to 1024. 
\subsubsection{Sentence-BERT.}
Afterwards, \verb|all-MiniLM-L6-v2|, the pre-trained model of Sentence-BERT \cite{Sentence-BERT2019}, is used to encode the surrounding text, to capture semantic features of the formula.

\subsection{Rank and Retrieval}\noindent
In order to improve retrieval performance, we adopt two-stage ranking. In the first stage, structural similarities between a query formula and each candidate formula in the collection are calculated based on their structural embeddings. We select the top 500{,}000 candidates as the initial retrieval results to reduce computation complexity in the second stage, in which semantic similarities between the query and each selected candidate are calculated and fused with their structural similarities through a weighted scheme. Specifically, let $f_q$ and $f_c$ denote the structural embeddings of the query formula and a candidate formula, and $t_q$ and $t_c$ denote the semantic embeddings of their surrounding text. The structural and semantic similarities are computed as follows: 
\begin{equation}
S_\text{structural}(q, c) = \cos(f_q, f_c);
\end{equation}
\begin{equation}
S_\text{semantic}(q, c) = \cos(t_q, t_c).
\end{equation}
Finally, a weighted similarity score is calculated as follows:
\begin{equation}
S_\text{final}(q, c) = \lambda \cdot S_\text{structural}(q, c) + (1 - \lambda) \cdot S_\text{semantic}(q, c),
\end{equation}
where $\lambda\in[0,1]$ is a hyperparameter controlling the similarity balance between formula structure and contextual semantics. The top 1{,}000 most similar candidates are selected based on $S_\text{final}$, as the final retrieval results.

\section{Experiments}
\subsection{Experimental Setup}
\subsubsection{Dataset.}
Experiments are conducted on the ARQMath-3 formula retrieval task\footnote{\url{https://www.cs.rit.edu/~dprl/ARQMath}}, which contains 76 queries and 28,320,920 formulas from 2,466,080 posts on Math Stack Exchange. For training we use 16,080,179 formulas, which are not in comments and with more than two nodes in their OPG representations, to reduce training time while ensuring data quality.
\subsubsection{Evaluation Metrics.}
In retrieval and recommendation systems, $P'@k$ and $nDCG'@k$ are commonly used to evaluate the quality of the top $k$ retrieved items \cite{ndcg2008}. Typically, we focus on $P^\prime@k$ for $k$=\{5, 10\} and $nDCG^\prime@k$ for $k$=10 after removing unjudged items from the final retrieval results. Following the ARQMath evaluation protocol, to calculate $P^\prime@k$, we treat only high and medium ratings as relevant. All metrics are calculated after formula instance deduplication by using ARQMath identifiers for visually distinct groupings.
\subsubsection{Hardware and Hyperparameters.}
The experiments are run on a server with two NVIDIA RTX 4090 GPUs (24GB each), 32 vCPUs, and 240 GB RAM. The StructEmb model are trained for 25 epochs with Adam optimization, embedding dimension of 400, batch size of 2560, learning rate of 1e-4, the number of GNN layers of 2, substructure substitution probabilities $p_{1}$=0.3, $p_{2}$=0.005 and $p_{3}$=0.002, attribute masking rate of 0.01 in the augmentation, and fusion hyperparameter $\lambda$=0.5. As for node embeddings, we firstly discard the node labels whose frequencies are less than 11 in all 175749 node labels of formulas, and then initialize the left 11868 node labels with random initialization. Finally, the node embeddings are pooled to obtain the initialized formula embeddings. 
\subsection{Experimental Results}\noindent
\balance
To evaluate SSEmb's utility for formula retrieval, we compare our results with the baseline Tangent-S \cite{davila2017layout} and representative runs submitted by participating teams in ARQMath-3. Specifically, for matching-based systems, including Tangent-S, Approach0 \cite{Approach0-ARQMath}, XY-Phoc \cite{XYPhoc-ARQMath}, MathDowsers \cite{MathDowsers-ARQMath} and JU\_NITS \cite{JU_NITS-ARQMath}, we report  results against their best-performing runs. For the embedding-based system DPRL \cite{DPRL-ARQMath}, we compare with its highest run, TangentCFT2ED, and its second-best, TangentCFT2, as shown in Table \ref{Tab1}. Among them Approach0 is a manual run that included human intervention. TangentCFT2 employs the TangentCFT \cite{Tangent-CFT2019} embedding-based model and combines results from SLT and OPT using score-weighted Reciprocal Rank Fusion (RRF) \cite{RRF2009}, while TangentCFT2ED further re-ranks the top results based on edit distance. SSEmb surpasses the best performance among the embedding-based systems by over 5 percentage points and also ranks highest among the automatically executed systems, approaching the best level of matching-based system Approach0.
\begin{table}[h]
  \caption{Performance of different methods. *Approach0 is a manual run while others are automated. \textsuperscript{+}SSEmb uses both formulas and surrounding text as input sources while others use only formulas.}
  \label{Tab1}
  \begin{tabular}{c|c|ccc}
    \toprule
    Type & Methods & $nDCG^\prime@10$ & $P^\prime@5$ & $P^\prime@10$ \\
    \hline
    \multirow{5}{*}{\makecell{Matching\\based}} 
                            & JU\_NITS     & 0.1648 & 0.1395 & 0.1250 \\
                            & Tangent-S    & 0.5765 & 0.5579 & 0.5105 \\
                            & MathDowsers  & 0.6081 & 0.6289 & 0.5487 \\
                            & XY-Phoc       & 0.6382 & 0.6316 & 0.5632 \\
                            & *Approach0    & {\bfseries 0.7511} & {\bfseries 0.7632} & {\bfseries 0.6882} \\
    \hline
    \multirow{3}{*}{\makecell{Embedding\\based}}  
                            & TangentCFT2  & 0.6211 & 0.6289 & 0.5342  \\
                            & TangentCFT2ED& 0.6868 & 0.7026 & 0.6105  \\
                            & \textsuperscript{+}SSEmb        & {\bfseries 0.7343} & {\bfseries 0.7632} & {\bfseries 0.6803} \\
    \bottomrule
  \end{tabular}
\end{table}
\subsection{Ablation Study}\noindent
We conduct ablation studies to investigate the necessity of the SemEmb module and the impacts of different graph data augmentations in the StructEmb module. The ablated results are reported in Table \ref{Tab2}.
\subsubsection{Necessity of the SemEmb Module.}Only using the StructEmb module from our SSEmb framework leads to a various decrease in $nDCG^\prime@10$, $P^\prime@5$ and $P^\prime@10$. This indicates that the contextual semantics encoded by the SemEmb module have significant impacts on the overall retrieval performance of SSEmb.

\subsubsection{Impacts of Augmentations in the StructEmb Module.} Graph data augmentations used in SSEmb consist of attribute masking and substructure substitution, which are perturbations on attributes and structures respectively. To further assess the impacts of different augmentations, we conduct three contrastive experiments using alternative combinations of graph data augmentations: (1) Substructure substitution; (2) Attribute masking+node dropping+edge perturbation; (3) Attribute masking. The experimental results show that combining attribute masking and substructure substitution yields the most effective performance. Notably, using substructure substitution alone also achieves competitive results, highlighting its effectiveness. Furthermore, under appropriate graph data augmentations, the StructEmb module, even when used independently, outperforms TangentCFT2 which also relies solely on formula embeddings. This underscores the potential of the StructEmb module in structure-aware formula encoding.
\begin{table}[h]
  \caption{Ablation study on SSEmb. \textsuperscript{+}SSEmb uses both formulas and surrounding text as input sources while others use only formulas.}
  \label{Tab2}
  \begin{tabular}{l|ccc}
    \toprule
    Methods  & $nDCG^\prime@10$ & $P^\prime@5$ & $P^\prime@10$ \\
    \hline
    \textsuperscript{+}SSEmb           & 0.7343 & 0.7632 & 0.6803  \\
    \hline
    - StructEmb   & 0.6762 & 0.6974 & 0.6211  \\
    \hline
    - StructEmb-subs               & 0.6744 & 0.6974 & 0.6013  \\
    - StructEmb-attr\_node\_edge   & 0.6298 & 0.6421 & 0.5658  \\
    - StructEmb-attr               & 0.5898 & 0.6329 & 0.5068  \\
    \bottomrule
  \end{tabular}
\end{table}
\subsection{Combining SSEmb's run with others}\noindent
To further improve the retrieval performance, we combine the highest original run from each participating team and SSEmb's run by RRF. As shown in Table \ref{Tab3},  the performance metrics of all the other methods are improved when combined with SSEmb. Among them, combining Approach0 with SSEmb outperforms the state-of-the-art system Approach0, and furthermore outperforms combining the original run of Approach0 with DPRL. This also shows that SSEmb performs better than the original embedding-based system DPRL.
\begin{table}[h]
  \caption{Performance of hybrid methods.}
  \label{Tab3}
  \centering
  \begin{tabular}{l|cc|cc}
    \toprule
    \multirow{2}{*}{\makecell{Methods}}&\multicolumn{2}{c|}{$nDCG^\prime@10$}&\multicolumn{2}{c}{$P^\prime@10$}\\
    &Original&RRF&Original&RRF\\
    \midrule
    Approach0\textsubscript{+SSEmb}     & 0.7511 & {\bfseries 0.7837} & 0.6882 & {\bfseries 0.7158} \\
    DPRL\textsubscript{+SSEmb}          & 0.6868 & {\bfseries 0.7197} & 0.6105 & {\bfseries 0.6434} \\
    MathDowsers\textsubscript{+SSEmb}   & 0.6081 & {\bfseries 0.6983} & 0.5487 & {\bfseries 0.6434} \\
    Tangent-S\textsubscript{+SSEmb}     & 0.5765 & {\bfseries 0.7011} & 0.5105 & {\bfseries 0.6434} \\
    XY-Phoc\textsubscript{+SSEmb}        & 0.6382 & {\bfseries 0.7008} & 0.5632 & {\bfseries 0.6500} \\
    JU\_NITS\textsubscript{+SSEmb}      & 0.1648 & {\bfseries 0.4369} & 0.1250 & {\bfseries 0.3711} \\
    \midrule
    Approach0\textsubscript{+DPRL}      & 0.7511 & 0.7459 & 0.6882 & 0.6724 \\
    \bottomrule
  \end{tabular}
\end{table}
\section{Conclusion}\noindent
In this paper, we propose SSEmb, a joint structual and semantic embedding framework for formula retrieval. SSEmb adapts Graph Contrastive Learning with a novel substructure substitution approach for graph data augmentation to encode structural features, and leverages Sentence-BERT to capture contextual semantics. The structural and semantic similarities are fused in a weighted scheme to improve retrieval performance.  Experimental results on ARQMath-3 show that SSEmb significantly outperforms the existing embedding-based models, ranks highest among the automatically executed systems, and achieves state-of-the-art results when combined with Approach0. In the next stage, we aim to explore more advanced techniques for graph representation learning and context encoding and further explore the utility of SSEmb on ARQMath’s Answer Retrieval task.
\balance
\bibliographystyle{ACM-Reference-Format}
\bibliography{references}


\begin{thebibliography}{29}


\ifx \showCODEN    \undefined \def \showCODEN     #1{\unskip}     \fi
\ifx \showDOI      \undefined \def \showDOI       #1{#1}\fi
\ifx \showISBNx    \undefined \def \showISBNx     #1{\unskip}     \fi
\ifx \showISBNxiii \undefined \def \showISBNxiii  #1{\unskip}     \fi
\ifx \showISSN     \undefined \def \showISSN      #1{\unskip}     \fi
\ifx \showLCCN     \undefined \def \showLCCN      #1{\unskip}     \fi
\ifx \shownote     \undefined \def \shownote      #1{#1}          \fi
\ifx \showarticletitle \undefined \def \showarticletitle #1{#1}   \fi
\ifx \showURL      \undefined \def \showURL       {\relax}        \fi
\providecommand\bibfield[2]{#2}
\providecommand\bibinfo[2]{#2}
\providecommand\natexlab[1]{#1}
\providecommand\showeprint[2][]{arXiv:#2}

\bibitem[Cormack et~al\mbox{.}(2009)]%
        {RRF2009}
\bibfield{author}{\bibinfo{person}{Gordon~V. Cormack}, \bibinfo{person}{Charles L~A Clarke}, {and} \bibinfo{person}{Stefan Buettcher}.} \bibinfo{year}{2009}\natexlab{}.
\newblock \showarticletitle{Reciprocal Rank Fusion Outperforms Condorcet and Individual Rank Learning Methods}. In \bibinfo{booktitle}{\emph{Proceedings of the 32nd International ACM SIGIR Conference on Research and Development in Information Retrieval (SIGIR)}}. \bibinfo{pages}{758–759}.
\newblock


\bibitem[Dai et~al\mbox{.}(2020)]%
        {Dai2020}
\bibfield{author}{\bibinfo{person}{Yifan Dai}, \bibinfo{person}{Liangyu Chen}, {and} \bibinfo{person}{Zihan Zhang}.} \bibinfo{year}{2020}\natexlab{}.
\newblock \showarticletitle{An N-ary Tree-based Model for Similarity Evaluation on Mathematical Formulae}. In \bibinfo{booktitle}{\emph{Proceedings of the 2020 IEEE International Conference on Systems, Man, and Cybernetics (SMC)}}. \bibinfo{pages}{2578--2584}.
\newblock


\bibitem[Davila and Zanibbi(2017)]%
        {davila2017layout}
\bibfield{author}{\bibinfo{person}{Kenny Davila} {and} \bibinfo{person}{Richard Zanibbi}.} \bibinfo{year}{2017}\natexlab{}.
\newblock \showarticletitle{Layout and Semantics: Combining Representations for Mathematical Formula Search}. In \bibinfo{booktitle}{\emph{Proceedings of the 40th International ACM SIGIR Conference on Research and Development in Information Retrieval (SIGIR)}}. \bibinfo{pages}{1165--1168}.
\newblock


\bibitem[Davila et~al\mbox{.}(2016)]%
        {davila2016tangent3}
\bibfield{author}{\bibinfo{person}{Kenny Davila}, \bibinfo{person}{Richard Zanibbi}, \bibinfo{person}{Andrew Kane}, {and} \bibinfo{person}{Frank~Wm. Tompa}.} \bibinfo{year}{2016}\natexlab{}.
\newblock \showarticletitle{Tangent-3 at the NTCIR-12 MathIR Task}. In \bibinfo{booktitle}{\emph{Proceedings of the 12th NTCIR Conference on Evaluation of Information Access Technologies (NTCIR)}}. \bibinfo{pages}{338--345}.
\newblock


\bibitem[Gao et~al\mbox{.}(2017)]%
        {gao2017preliminaryexplorationformulaembedding}
\bibfield{author}{\bibinfo{person}{Liangcai Gao}, \bibinfo{person}{Zhuoren Jiang}, \bibinfo{person}{Yue Yin}, \bibinfo{person}{Ke Yuan}, \bibinfo{person}{Zuoyu Yan}, {and} \bibinfo{person}{Zhi Tang}.} \bibinfo{year}{2017}\natexlab{}.
\newblock \showarticletitle{Preliminary Exploration of Formula Embedding for Mathematical Information Retrieval: Can Mathematical Formulae be Embedded like A Natural Language?}
\newblock \bibinfo{journal}{\emph{arXiv preprint arXiv:1707.05154}} (\bibinfo{year}{2017}).
\newblock


\bibitem[Hu et~al\mbox{.}(2013)]%
        {Wikimirs2013}
\bibfield{author}{\bibinfo{person}{Xuan Hu}, \bibinfo{person}{Liangcai Gao}, \bibinfo{person}{Xiaoyan Lin}, \bibinfo{person}{Zhi Tang}, \bibinfo{person}{Xiaofan Lin}, {and} \bibinfo{person}{Josef~B. Baker}.} \bibinfo{year}{2013}\natexlab{}.
\newblock \showarticletitle{WikiMirs: A Mathematical Information Retrieval System for Wikipedia}. In \bibinfo{booktitle}{\emph{Proceedings of the 13th ACM/IEEE-CS Joint Conference on Digital Libraries (JCDL)}}. \bibinfo{pages}{11–20}.
\newblock


\bibitem[Kane et~al\mbox{.}(2022)]%
        {MathDowsers-ARQMath}
\bibfield{author}{\bibinfo{person}{Andrew Kane}, \bibinfo{person}{Yin Ng}, {and} \bibinfo{person}{Frank Tompa}.} \bibinfo{year}{2022}\natexlab{}.
\newblock \showarticletitle{Dowsing for Answers to Math Questions: Doing Better With Less}. In \bibinfo{booktitle}{\emph{Proceedings of the 13th Conference and Labs of the Evaluation Forum (CLEF)}}. \bibinfo{pages}{40--62}.
\newblock


\bibitem[Kristianto et~al\mbox{.}(2016)]%
        {MCAT2016}
\bibfield{author}{\bibinfo{person}{Giovanni~Yoko Kristianto}, \bibinfo{person}{Goran Topic}, {and} \bibinfo{person}{Akiko Aizawa}.} \bibinfo{year}{2016}\natexlab{}.
\newblock \showarticletitle{MCAT Math Retrieval System for NTCIR-12 MathIR Task}. In \bibinfo{booktitle}{\emph{Proceedings of the 12th NTCIR Conference on Evaluation of Information Access Technologies (NTCIR)}}. \bibinfo{pages}{323–330}.
\newblock


\bibitem[Krstovski and Blei(2018)]%
        {EquationEmbeddings2018}
\bibfield{author}{\bibinfo{person}{Kriste Krstovski} {and} \bibinfo{person}{David~M. Blei}.} \bibinfo{year}{2018}\natexlab{}.
\newblock \showarticletitle{Equation Embeddings}.
\newblock \bibinfo{journal}{\emph{arXiv preprint arXiv:1803.09123}} (\bibinfo{year}{2018}).
\newblock


\bibitem[Kumar et~al\mbox{.}(2012)]%
        {kumar2012structure}
\bibfield{author}{\bibinfo{person}{P.~Pavan Kumar}, \bibinfo{person}{Arun Agarwal}, {and} \bibinfo{person}{Chakravarthy Bhagvati}.} \bibinfo{year}{2012}\natexlab{}.
\newblock \showarticletitle{A Structure Based Approach for Mathematical Expression Retrieval}. In \bibinfo{booktitle}{\emph{Proceedings of the 6th Multi-Disciplinary International Workshop on Artificial Intelligence (MIWAI)}}. \bibinfo{pages}{23--34}.
\newblock


\bibitem[Langsenkamp et~al\mbox{.}(2022)]%
        {XYPhoc-ARQMath}
\bibfield{author}{\bibinfo{person}{Matt Langsenkamp}, \bibinfo{person}{Bryan Amador}, {and} \bibinfo{person}{Richard Zanibbi}.} \bibinfo{year}{2022}\natexlab{}.
\newblock \showarticletitle{Expanding Spatial Regions and Incorporating IDF for PHOC-Based Math Formula Retrieval at ARQMath-3}. In \bibinfo{booktitle}{\emph{Proceedings of the 13th Conference and Labs of the Evaluation Forum (CLEF)}}. \bibinfo{pages}{63--82}.
\newblock


\bibitem[Mansouri et~al\mbox{.}(2022a)]%
        {10.1007/978-3-031-13643-6_20}
\bibfield{author}{\bibinfo{person}{Behrooz Mansouri}, \bibinfo{person}{V{\'i}t Novotn{\'y}}, \bibinfo{person}{Anurag Agarwal}, \bibinfo{person}{Douglas~W. Oard}, {and} \bibinfo{person}{Richard Zanibbi}.} \bibinfo{year}{2022}\natexlab{a}.
\newblock \showarticletitle{Overview of ARQMath-3 (2022): Third CLEF Lab on Answer Retrieval for Questions on Math}. In \bibinfo{booktitle}{\emph{Proceedings of the 13th Conference and Labs of the Evaluation Forum (CLEF)}}. \bibinfo{pages}{1--27}.
\newblock


\bibitem[Mansouri et~al\mbox{.}(2022b)]%
        {DPRL-ARQMath}
\bibfield{author}{\bibinfo{person}{Behrooz Mansouri}, \bibinfo{person}{Douglas Oard}, {and} \bibinfo{person}{Richard Zanibbi}.} \bibinfo{year}{2022}\natexlab{b}.
\newblock \showarticletitle{DPRL Systems in the CLEF 2022 ARQMath Lab: Introducing MathAMR for Math-Aware Search}. In \bibinfo{booktitle}{\emph{Proceedings of the 13th Conference and Labs of the Evaluation Forum (CLEF)}}. \bibinfo{pages}{83--103}.
\newblock


\bibitem[Mansouri et~al\mbox{.}(2022c)]%
        {MathAMR2022}
\bibfield{author}{\bibinfo{person}{Behrooz Mansouri}, \bibinfo{person}{Douglas~W. Oard}, {and} \bibinfo{person}{Richard Zanibbi}.} \bibinfo{year}{2022}\natexlab{c}.
\newblock \showarticletitle{Contextualized Formula Search Using Math Abstract Meaning Representation}. In \bibinfo{booktitle}{\emph{Proceedings of the 31st ACM International Conference on Information \& Knowledge Management (CIKM)}}. \bibinfo{pages}{4329–4333}.
\newblock


\bibitem[Mansouri et~al\mbox{.}(2019)]%
        {Tangent-CFT2019}
\bibfield{author}{\bibinfo{person}{Behrooz Mansouri}, \bibinfo{person}{Shaurya Rohatgi}, \bibinfo{person}{Douglas~W. Oard}, \bibinfo{person}{Jian Wu}, \bibinfo{person}{C.~Lee Giles}, {and} \bibinfo{person}{Richard Zanibbi}.} \bibinfo{year}{2019}\natexlab{}.
\newblock \showarticletitle{Tangent-CFT: An Embedding Model for Mathematical Formulas}. In \bibinfo{booktitle}{\emph{Proceedings of the 42nd International ACM SIGIR Conference on Research and Development in Information Retrieval (SIGIR)}}. \bibinfo{pages}{11–18}.
\newblock


\bibitem[Miller and Youssef(2003)]%
        {miller2003technical}
\bibfield{author}{\bibinfo{person}{Bruce~R. Miller} {and} \bibinfo{person}{Abdou~S. Youssef}.} \bibinfo{year}{2003}\natexlab{}.
\newblock \showarticletitle{Technical Aspects of the Digital Library of Mathematical Functions}.
\newblock \bibinfo{journal}{\emph{Annals of Mathematics and Artificial Intelligence}} \bibinfo{volume}{38}, \bibinfo{number}{1--3} (\bibinfo{year}{2003}), \bibinfo{pages}{121--136}.
\newblock


\bibitem[Mišutka and Galamboš(2008)]%
        {misutka2008extending}
\bibfield{author}{\bibinfo{person}{Jozef Mišutka} {and} \bibinfo{person}{Leo Galamboš}.} \bibinfo{year}{2008}\natexlab{}.
\newblock \showarticletitle{Extending Full Text Search Engine for Mathematical Content}.
\newblock \bibinfo{journal}{\emph{Towards Digital Mathematics Library}} (\bibinfo{year}{2008}), \bibinfo{pages}{55--67}.
\newblock


\bibitem[Oord et~al\mbox{.}(2018)]%
        {InfoNCELoss2018}
\bibfield{author}{\bibinfo{person}{Aaron van~den Oord}, \bibinfo{person}{Yazhe Li}, {and} \bibinfo{person}{Oriol Vinyals}.} \bibinfo{year}{2018}\natexlab{}.
\newblock \showarticletitle{Representation Learning with Contrastive Predictive Coding}.
\newblock \bibinfo{journal}{\emph{arXiv preprint arXiv:1807.03748}} (\bibinfo{year}{2018}).
\newblock


\bibitem[Reimers and Gurevych(2019)]%
        {Sentence-BERT2019}
\bibfield{author}{\bibinfo{person}{Nils Reimers} {and} \bibinfo{person}{Iryna Gurevych}.} \bibinfo{year}{2019}\natexlab{}.
\newblock \showarticletitle{Sentence-{BERT}: Sentence Embeddings using {S}iamese {BERT}-Networks}. In \bibinfo{booktitle}{\emph{Proceedings of the 2019 Conference on Empirical Methods in Natural Language Processing and the 9th International Joint Conference on Natural Language Processing (EMNLP-IJCNLP)}}. \bibinfo{pages}{3982--3992}.
\newblock


\bibitem[Sakai and Kando(2008)]%
        {ndcg2008}
\bibfield{author}{\bibinfo{person}{Tetsuya Sakai} {and} \bibinfo{person}{Noriko Kando}.} \bibinfo{year}{2008}\natexlab{}.
\newblock \showarticletitle{On Information Retrieval Metrics Designed for Evaluation with Incomplete Relevance Assessments}.
\newblock \bibinfo{journal}{\emph{Information Retrieval}}  \bibinfo{volume}{11} (\bibinfo{year}{2008}), \bibinfo{pages}{447--470}.
\newblock


\bibitem[Sarkar et~al\mbox{.}(2022)]%
        {JU_NITS-ARQMath}
\bibfield{author}{\bibinfo{person}{Sandip Sarkar}, \bibinfo{person}{Dipankar Das}, \bibinfo{person}{Partha Pakray}, {and} \bibinfo{person}{David Pinto}.} \bibinfo{year}{2022}\natexlab{}.
\newblock \showarticletitle{Formula Retrieval Using Structural Similarity}. In \bibinfo{booktitle}{\emph{Proceedings of the 13th Conference and Labs of the Evaluation Forum (CLEF)}}. \bibinfo{pages}{138--146}.
\newblock


\bibitem[Song and Chen(2021)]%
        {Song2021}
\bibfield{author}{\bibinfo{person}{Yujin Song} {and} \bibinfo{person}{Xiaoyu Chen}.} \bibinfo{year}{2021}\natexlab{}.
\newblock \showarticletitle{Searching for Mathematical Formulas Based on Graph Representation Learning}. In \bibinfo{booktitle}{\emph{Proceedings of the 14th Conference on Intelligent Computer Mathematics (CICM)}}. \bibinfo{pages}{137--152}.
\newblock


\bibitem[Thanda et~al\mbox{.}(2016)]%
        {Thanda2016}
\bibfield{author}{\bibinfo{person}{Abhinav Thanda}, \bibinfo{person}{Ankit Agarwal}, \bibinfo{person}{Kushal Singla}, \bibinfo{person}{Aditya Prakash}, {and} \bibinfo{person}{Abhishek Gupta}.} \bibinfo{year}{2016}\natexlab{}.
\newblock \showarticletitle{A Document Retrieval System for Math Queries}. In \bibinfo{booktitle}{\emph{Proceedings of the 12th NTCIR Conference on Evaluation of Information Access Technologies (NTCIR)}}. \bibinfo{pages}{346–353}.
\newblock


\bibitem[Wang and Chen(2025)]%
        {GCL-MIR}
\bibfield{author}{\bibinfo{person}{Pei-Syuan Wang} {and} \bibinfo{person}{Hung-Hsuan Chen}.} \bibinfo{year}{2025}\natexlab{}.
\newblock \showarticletitle{The Effectiveness of Graph Contrastive Learning on Mathematical Information Retrieval}. In \bibinfo{booktitle}{\emph{Proceedings of the 1st International Workshop on Graph-Based Approaches in Information Retrieval (IRonGraphs)}}. \bibinfo{pages}{60--72}.
\newblock


\bibitem[Wang et~al\mbox{.}(2015)]%
        {WikiMirs3.02015}
\bibfield{author}{\bibinfo{person}{Yuehan Wang}, \bibinfo{person}{Liangcai Gao}, \bibinfo{person}{Simeng Wang}, \bibinfo{person}{Zhi Tang}, \bibinfo{person}{Xiaozhong Liu}, {and} \bibinfo{person}{Ke Yuan}.} \bibinfo{year}{2015}\natexlab{}.
\newblock \showarticletitle{WikiMirs 3.0: A Hybrid MIR System Based on the Context, Structure and Importance of Formulae in a Document}. In \bibinfo{booktitle}{\emph{Proceedings of the 15th ACM/IEEE-CS Joint Conference on Digital Libraries (JCDL)}}. \bibinfo{pages}{173–182}.
\newblock


\bibitem[Yasunaga and Lafferty(2019)]%
        {TopicEq2019}
\bibfield{author}{\bibinfo{person}{Michihiro Yasunaga} {and} \bibinfo{person}{John~D. Lafferty}.} \bibinfo{year}{2019}\natexlab{}.
\newblock \showarticletitle{TopicEq: A Joint Topic and Mathematical Equation Model for Scientific Texts}. In \bibinfo{booktitle}{\emph{Proceedings of the 33rd AAAI Conference on Artificial Intelligence (AAAI)}}. \bibinfo{pages}{7394–7401}.
\newblock


\bibitem[You et~al\mbox{.}(2020)]%
        {GraphCL2020}
\bibfield{author}{\bibinfo{person}{Yuning You}, \bibinfo{person}{Tianlong Chen}, \bibinfo{person}{Yongduo Sui}, \bibinfo{person}{Ting Chen}, \bibinfo{person}{Zhangyang Wang}, {and} \bibinfo{person}{Yang Shen}.} \bibinfo{year}{2020}\natexlab{}.
\newblock \showarticletitle{Graph Contrastive Learning with Augmentations}. In \bibinfo{booktitle}{\emph{Proceedings of the 34th International Conference on Neural Information Processing Systems (NeurIPS)}}. \bibinfo{pages}{5812--5823}.
\newblock


\bibitem[Zhong et~al\mbox{.}(2022)]%
        {Approach0-ARQMath}
\bibfield{author}{\bibinfo{person}{Wei Zhong}, \bibinfo{person}{Yuqing Xie}, {and} \bibinfo{person}{Jimmy Lin}.} \bibinfo{year}{2022}\natexlab{}.
\newblock \showarticletitle{Applying Structural and Dense Semantic Matching for the ARQMath Lab 2022, ClEF}. In \bibinfo{booktitle}{\emph{Proceedings of the 13th Conference and Labs of the Evaluation Forum (CLEF)}}. \bibinfo{pages}{147--170}.
\newblock


\bibitem[Zhong and Zanibbi(2019)]%
        {zhong2019structural}
\bibfield{author}{\bibinfo{person}{Wei Zhong} {and} \bibinfo{person}{Richard Zanibbi}.} \bibinfo{year}{2019}\natexlab{}.
\newblock \showarticletitle{Structural Similarity Search for Formulas Using Leaf-Root Paths in Operator Subtrees}. In \bibinfo{booktitle}{\emph{Proceedings of the 41st European Conference on Information Retrieval (ECIR)}}. \bibinfo{pages}{116--129}.
\newblock


\end{thebibliography}
\end{document}